\documentclass[
  reprint,            
  amsmath,amssymb,    
  aps,                
  prx,                
  floatfix,            
  superscriptaddress
]{revtex4-2}
\usepackage{graphicx} 
\usepackage{braket} 
\newcommand{\norm}[1]{\left\lVert#1\right\rVert}
\usepackage{xcolor}

\begin{document}

\title{\bf Performance of Krotov, PRONTO and PINN for optimal control of quantum gates}

\author{M. D. Jim\'{e}nez}
\email{martin.jimenez.388@mi.unc.edu.ar}
\affiliation{Departamento de F\'isica, Universidade Federal de S\~ao Carlos (UFSCar)\\ S\~ao Carlos, SP 13565-905, Brazil}
\author{M. D. Forlevesi} 
\email{murilo.deliberali@unesp.br}
\affiliation{Departamento de F\'{i}sica, Universidade Estadual Paulista, Rio Claro, SP, Brazil.}
\author{E. F. de Lima}
\email{eflima@ufscar.br}
\author{L. K. Castelano}
\email{lkcastelano@ufscar.br}
\affiliation{Departamento de F\'isica, Universidade Federal de S\~ao Carlos (UFSCar)\\ S\~ao Carlos, SP 13565-905, Brazil}
\date{\today}

\begin{abstract}
Achieving scalable quantum computing demands high-fidelity operations capable of mitigating population leakage into non-computational states. Physics-Informed Neural Networks (PINNs) have recently emerged as a powerful paradigm to unify quantum hardware characterization (inverse problems) and pulse engineering (direct problems), laying the foundational architecture for autonomous quantum processors. However, standard PINN frameworks face severe numerical bottlenecks, such as spectral bias, when attempting to simultaneously solve highly oscillatory multi-level dynamics and optimize continuous control fields under strict global phase constraints. In this work, we propose an enhanced PINN scheme for quantum optimal control (PINNQOC) that circumvents these limitations by incorporating Fourier feature embeddings, dynamic epoch normalization, and an informed pre-training routine. To rigorously evaluate its performance, we systematically benchmark our framework against two premier continuous control solvers: the first-order Krotov method and the second-order Projection Operator Newton Method for Trajectory Optimization (PRONTO). These techniques are applied to implement multiple quantum gates on a truncated three-level fluxonium qubit and a four-level Nitrogen-Vacancy center coupled to a Carbon-13 nuclear spin. Our advanced PINNQOC approach successfully suppresses population leakage while achieving gate fidelities exceeding 99.9$\%$, matching the efficacy of traditional solvers. Finally, we provide a comprehensive analysis of computational times, iteration efficiency, and mean leakage, highlighting the distinct trade-offs and avenues for embedding physics-guided machine learning into automated quantum hardware pipelines.
\end{abstract}
\maketitle

\date{\today}




\pagebreak
\section{Introduction}
\label{sec:introduction}

Achieving the promise of quantum computing for practical applications requires the precise ability to fine-tune and control multi-qubit systems \cite{Nielsen2000}. However, multiple experimental challenges currently prevent state-of-the-art quantum processors from executing quantum operations with the high fidelities required for interesting computing tasks. Among the primary sources of error in quantum information processing are the coupling of the system to its surrounding environment, leading to open quantum system dynamics~\cite{breuer2002theory}, crosstalk between adjacent qubits~\cite{gambetta2012characterization,mckay2019three}, and spectral noise within the external driving fields~\cite{PhysRevA.107.042611} . Consequently, any realistic external control strategy must simultaneously mitigate these diverse decoherence channels while ensuring that the system reliably performs the target quantum operations.
A particularly critical challenge in this domain is the leakage of population from the computational subspace into higher-lying, non-computational energy levels \cite{leak1,leak2,leak3,leak4}. In many physical implementations, such as superconducting circuits \cite{Babu2021} or solid-state defects \cite{delMoral2026}, qubits are modeled as truncated multi-level systems. The presence of these leaking levels significantly degrades the performance and fidelity of quantum gates. 

In the Quantum Optimal Control (QOC) literature, a wide variety of frameworks have been proposed to address these requirements~\cite{Glaser2015}. Traditional approaches often employ piecewise constant fields optimized through gradient-based algorithms such as GRAPE and CRAB \cite{KHANEJA2005296,PhysRevA.84.022326,PhysRevLett.106.190501}. Alternatively, methods like Krotov and the Projection Operator Newton Method for Trajectory Optimization (PRONTO) treat control fields as arbitrary continuous functions \cite{GoerzPackage,Goerz_2014,Shao2024,Fernandes_2023}.

Recently, Physics-Informed Neural Networks (PINNs)~\cite{RAISSI2019686} have emerged as a versatile framework capable of addressing a wide spectrum of quantum dynamics challenges, effectively bridging the gap between system characterization and operation~\cite{DeRyck_Mishra_2024}. On the one hand, the PINNverse methodology has been successfully established to solve the inverse problem, efficiently extracting underlying Hamiltonian parameters and identifying physical parameters from noisy experimental datasets ~\cite{DELIMA2026131548, PhysRevA.110.032607}. On the other hand, the direct problem of QOC, where continuous control fields and state dynamics are jointly modeled within the output layers of a NN, has shown great theoretical promise~\cite{Norambuena2023,PhysRevX.8.031086}. By framing both Hamiltonian learning and quantum control as loss-function minimization tasks constrained by physical laws, PINNs offer a unified computational ecosystem. This mathematical synergy is a crucial step toward fully autonomous quantum processors, enabling hardware characterization and in situ pulse optimization to be integrated into a single, closed-loop pipeline. Under this paradigm, as a physical platform suffers from parameter drift~\cite{PhysRevLett.106.180504} or environmental noise over time~\cite{carroll2022dynamicssuperconductingqubitrelaxation}, the network can dynamically update its Hamiltonian model (the inverse problem) and immediately adapt the control fields to maintain high-fidelity operations (the direct problem) without requiring external manual recalibration. Within this overarching framework, the same foundational machine-learning architecture utilized to precisely characterize a quantum platform via the PINNverse approach can be naturally extended to design high-fidelity quantum gates. Despite these promising developments, there remains a critical gap in the literature regarding a systematic comparison between machine-learning-based approaches and highly specialized QOC solvers.



Although recent exploratory efforts have compared PINN-based control with standard pulse engineering frameworks in low-dimensional systems \cite{Norambuena2023}, a comprehensive benchmark is still lacking. In this work, we systematically assess the performance of PINN for quantum optimal control (PINNQOC) against the well-established Krotov method~\cite{GoerzPackage}. Moreover, we use PRONTO, which is a second-order optimal control algorithm that solves the QOC problem~\cite{Shao2022,Shao2024}. Unlike first-order optimization methods, which update the control fields using only gradient information, PRONTO constructs a local second-order approximation of the optimization landscape and computes Newton search directions in the space of dynamically feasible trajectories. A key feature of the method is the introduction of a nonlinear projection operator that directly incorporates the system dynamics into the optimization procedure, ensuring that each optimization iteration satisfies the Schrödinger equation without the need for explicit Lagrange multipliers. Furthermore, we demonstrate that PINNQOC exhibits intrinsic numerical difficulties even for relatively simple systems. To overcome these limitations, we propose a new scheme that enables the efficient implementation of PINNQOC using a compact neural network architecture.

\section{Gate implementation as a control problem}
We are interested in solving the problem of QOC for the implementation of quantum gates. First, we have to solve the Schr\"odinger equation
\begin{equation}
i\frac{d|\psi(t)\rangle}{dt} =H(t)\ket{\psi(t)}=  \left( H_0+  H_c(t)\right)|\psi(t)\rangle,\label{eq:scrodinger}
\end{equation}
where $ H_0$ is a time-independent term, $ H_c(t)=\sum_{m=1}^M u_m(t) H_m$ denotes the time-dependent term, and $u_m(t)$ is the control field associated with the coupling terms described by $ H_m$. Throughout this paper, we always take $\hbar=1$. The control fields must be optimized to evolve a set of initial states $\ket{\Psi_0}=(\ket{\psi_1^0},...,\ket{\psi_N^0})^T$ to their respective target states $\ket{\Psi_{\text{tgt}}}=(\ket{\psi_1^{\text{tgt}}},...,\ket{\psi_N^{\text{tgt}}})^T$. When the goal is to optimize the controls to perform a quantum gate, $N$ is equal to the dimension of the state space related to the computational states. Furthermore, phases play an important role in the optimization functional, and we discuss two cases. The first case allows for an arbitrary global phase; the desired quantum gate $U^{\text{tgt}}$ obtained by the optimization is given by $e^{i\varphi}U^{\text{tgt}}$, where $\varphi$ is the global phase. The functional for this case can be written as
\begin{equation}
    J_{\varphi}\left[\ket{\Psi(T)} \right]=1-\frac{1}{N^2}\left|\langle\Psi_{\text{tgt}}|\Psi(T)\rangle\right|^2,\label{eq:Jphi}
\end{equation}
where $|\Psi(T)\rangle=(\ket{\psi_1(T)},\ldots,\ket{\psi_N(T)})^T$ denotes the final-time state corresponding to the initial state $|\Psi^0\rangle$. If the optimization requires the global phase to be set to zero $\varphi=0$, the functional becomes
\begin{equation}
    J_{0}\left[\ket{\Psi(T)} \right]=\frac{1}{N}\norm{|\Psi_{\text{tgt}}\rangle-|\Psi(T)\rangle}^2,
    \label{eq:gate_functional}
\end{equation}
where $\norm{|\Psi_a\rangle}^2=\langle\Psi_a\ket{\Psi_a}.$

\section{Krotov method with penalization}
Given the set of control fields represented by $\boldsymbol{u}=(u_1(t),u_2(t),\ldots,u_M(t))$, we want to find the total Hamiltonian $H(t)=H_0+H_c(t)$ from Eq.\eqref{eq:scrodinger} to produce the time evolution that satisfies $\ket{\Psi_0}\to \ket{\Psi(T)}=\ket{\Psi_{\text{tgt}}}$ at the final time of evolution $T$. In the Krotov method, some further constraints are typically required, such as having a low energy fluence of control fields. In addition, we can require some extra constraints related to the time-evolved states. A typical example is when we consider a multi-level system, but we are only interested in performing operations over the subspace spanned by the eigenstates corresponding to the two lowest levels, while keeping the occupancies of higher levels minimal throughout the evolution, effectively producing a qubit operation. In this work, we focus our attention on two constraints: enforcing smooth changes in the control fields during the optimization and minimizing the leakage to higher levels. This problem can be mathematically formulated as the search for control fields $\boldsymbol{u}$ and states $\ket{\Psi}$ that minimize the composite functional $J\left[\ket{\Psi} ,\boldsymbol{u}\right]$~\cite{GoerzPackage}, defined as follows
\begin{widetext}
\begin{equation}
\begin{split}
J\left[\ket{\Psi} ,\boldsymbol{u}\right] &=J_{0}\left[\ket{\Psi(T)}\right] +\int _{0}^{T} g_{a}(\boldsymbol{u} ,t) dt+\int _{0}^{T} g_{b}\left(\ket{\Psi(t)} ,t\right) dt,
\end{split}
\label{eq_func}
\end{equation}
\end{widetext}
where
\begin{equation}
    g_{a}(\boldsymbol{u} ,t)=\sum _{m=1}^{M}\frac{\lambda _m}{ S_m( t)}\left( u_m( t) -u_m^{( 0)}( t)\right)^{2},\nonumber
\end{equation}
and
\begin{equation}
    g_{b}\left(\ket{\Psi(t)},t\right)= q \sum _{j=1}^{N}\bra{\psi _j( t)}{P}\ket{\psi _j( t)}.\nonumber
\end{equation}

\noindent In the above equations, $S_m(t)$ are the envelope functions, $\lambda_m$ are step widths, $u_m^{(0)}$ are the guess control fields, $q$ is a weight factor, and ${P}$ is the projection operator over a subspace of states that we want to avoid populating during the time evolution. The functional $ g_{a}(\boldsymbol{u} ,t)$ ensures smooth updates for the control fields and $g_{b}\left(\ket{\Psi(t)} ,t\right)$ penalizes populations of higher levels. 
In addition, states and control fields are constrained to satisfy the Schr\"odinger equation \eqref{eq:scrodinger}, which can be imposed by including a Lagrange multiplier in the definition of the functional. Once we have built our augmented functional, we have to request the minimization condition $\delta J=0$. This leads to the following update rule from the {\it k}-th to the ({\it k+1})-th iteration for each control field
\begin{widetext}
\begin{eqnarray}
u_m^{( k+1)}( t) =u_m^{( k)}( t)+ \frac{S_m( t)}{\lambda _m} \ \sum _{j=1}^{N}\text{Im}\Big\langle\chi _j^{( k)}( t)\left|\frac{\partial {H}}{\partial u_m}\right|\psi _j^{( k+1)}( t)\Big\rangle,
\label{eq:krotov_update}
\end{eqnarray}
\end{widetext}
where the states $\ket{\psi _j^{( k)}( t)}$ are forward-evolved under the Hamiltonian ${H}^{(k)}(t)$ containing the fields of the {\it k}-th iteration, and the backward-evolved states $\ket{\chi _j^{(k)}( t)}$ satisfy the inhomogeneous equation
\begin{equation}
\frac{d\ket{\chi _j^{(k)}( t)}}{dt} =-i{H}^{(k)}(t)\ket{\chi _j^{(k)}( t)} +q{P}\ket{\psi _j^{(k)}( t)},
\label{eq:Krot_costates}
\end{equation}
with the conditions $\ket{\chi _j^{( k)}( T)} =\ket{\psi _j^{\text{tgt}}}$ for $j=1,\ldots,N$.
Note that the forward-evolved states $\ket{\psi _j^{( k)}( t)}$ are evolved using the control fields of the {\it k}-th iteration itself. To understand how this works, we need to dive deeper into the numerical implementation. First, we choose a finite time grid $(t_0,...,t_{N_t})$, where $N_t$ is the number of time steps, $t_0=0$ and $t_{N_t}=T$. The states and co-states will be evaluated over this time grid, while the control fields will be defined at its midpoints. Therefore, in the update rule Eq.~\eqref{eq:krotov_update}, the fields at the point $\tilde t_{n}=(t_n+t_{n+1})/2$ will be updated by evaluating the states in the second term of the right-hand side at $t=t_n$. In turn, the state $\ket{\psi _{j}^{( k)}(t_n)}$ had previously been propagated from $t_{n-1}$ to $t_n$ using the fields at $\tilde t_{n-1} =(t_{n-1}+t_n)/2$, which had been updated using the states at $t_{n-1}$, and so on. Even though Krotov's method assumes that the control fields are continuous functions, the numerical implementation requires some propagation method to perform the time evolution. It is common to use the rule $\ket{\psi _{j}^{( k)}(t_{n+1})}=e^{-iH^{(k)}(\tilde t_n)(t_{n+1}-t_n)}\ket{\psi _{j}^{( k)}(t_{n})}$ to forward propagate the states, assuming that the control fields are piece-wise constant in the time grid. But in our case, the same evolution method cannot be used to backward propagate the co-states, since the inhomogeneity in Eq. \eqref{eq:Krot_costates} makes this approximation invalid. Instead, we approximate the control fields by linear splines at the midpoints, linearly extrapolate the fields to the endpoints $t_0$ and $t_{N_t}$, and perform the evolution using the $4$-th order Runge-Kutta method.

\section{Projection operator
Newton method for trajectory optimization}


PRONTO is a second-order optimization framework that enforces the system dynamics through a trajectory projection operator while exploiting Newton iterations to accelerate convergence. In this formulation, each optimization step is obtained by solving a local linear-quadratic optimal control problem derived from a second-order approximation of the projected cost functional. The present section summarizes the main ingredients of the PRONTO algorithm required for the development of this work. For a complete description of its theoretical formulation and numerical implementation, the reader is referred to Refs.~\cite{Shao2022,Shao2024}.

For convenience, we introduce a compact notation for the quantum gate synthesis problem. The propagated states are assembled into the stacked state vector
\begin{equation}
|\Psi(t)\rangle=\left(|\psi_1(t)\rangle,|\psi_2(t)\rangle,\ldots,|\psi_N(t)\rangle\right)^T,
\end{equation}
together with the block-diagonal Hamiltonian
\begin{equation}
H_N(t)=I_N\otimes H(t),
\end{equation}

where \(I_N\) is the identity matrix of dimension \(N\) and \(\otimes\) denotes the Kronecker product. 
To solve the control problem we start by transforming the complex state vector $\ket{\Psi(t)}\in \mathbb{C}^n$ into a larger real vector $x(t)\in\mathbb{R}^{2n}$ through the bijective mapping
\begin{equation}
  x(t)=
\begin{bmatrix}
\mathrm{Re}[\ket{\Psi(t)}]\\
\mathrm{Im}[\ket{\Psi(t)}]
\end{bmatrix}.
\end{equation}
The Schrödinger equation \eqref{eq:scrodinger} can also be rewritten as the following real-valued ordinary differential equation,

\begin{equation}
\dot{x}(t)=\mathcal{H}_N(t) \, x(t),
\label{eq:pronto_real}
\end{equation}
\noindent where the real-valued matrix $\mathcal{H}_N$ is given by

\begin{equation}
\mathcal{H}_N=
\begin{bmatrix}
\operatorname{Re}(-iH_N) &
-\operatorname{Im}(-iH_N) \\
\operatorname{Im}(-iH_N) &
\operatorname{Re}(-iH_N)
\end{bmatrix}.
\end{equation}

The mapping from $\mathbb{C}^n$ to $\mathbb{R}^{2n}$ is employed to transform the cost functional into its real-valued equivalent form, following Ref.~\cite{Shao2022}. Throughout this work, the terminal cost functional $J_0$ and the state cost $g_b$ are kept unchanged from the Krotov formulation. The only modification concerns the control cost, which is taken as
\begin{equation}
g_a(\boldsymbol{u},t)=\frac{1}{2}\boldsymbol{u}^{T}(t)\,\mathcal{R}(t)\boldsymbol{u}(t),
\label{eq:lu_pronto}
\end{equation}
where $\mathcal{R}(t)$ is a positive-definite weighting matrix that penalizes the control fluence. Unlike the Krotov formulation, which penalizes variations with respect to a reference control, the PRONTO framework  penalizes the control energy directly.




The quantum optimal control problem is also equivalent to the Banach space optimization problem,
\begin{equation}
\min_{\xi \in \mathcal{T}} h(\xi),
\label{eq:banach_problem}
\end{equation}
where $\xi = \bigl[x(t),\,\boldsymbol{u}(t)\bigr]$ is a state-and-input trajectory pair,

\begin{align}
   h(\xi) = J_0(x(T), \boldsymbol{u})+\int_{0}^{T}g_a\left(\boldsymbol{u},t\right)\,dt \nonumber\\
+\int_{0}^{T}g_b\left(\ket{\psi},t\right)\,dt
\label{eq:cost_functional} 
\end{align}
is the cost functional, and
\begin{equation}
\mathcal{T} =\left\{\xi\;\middle|\;\dot{x}(t)=\mathcal{H}_N(t) \,x(t),\;x(0)=x_0\right\}
\label{eq:trajectory_manifold}
\end{equation}
is the trajectory manifold.

The main idea of PRONTO is to eliminate the dynamic constraints by introducing a trajectory projection operator. Instead of directly optimizing admissible state-input trajectories, the algorithm optimizes arbitrary state-input curves and projects them onto the trajectory manifold at every iteration. Consequently, the original constrained optimization problem \eqref{eq:banach_problem} is reformulated as the unconstrained optimization problem
\begin{equation}
\min g(\eta),
\label{eq:unconstrained}
\end{equation}
where $\eta = \bigl[\alpha(t),\,\mu(t)\bigr]$  is a pair of state-and-input curves that may not belong to the trajectory manifold $\mathcal{T}$. This is done introducing a projection operator $\mathcal{P}(\eta)$ such that, 
\begin{equation}
g(\eta) = h[\mathcal{P}(\eta)],
\end{equation}
where  $\mathcal{P}(\eta)$ projects any \(\eta\) onto the
manifold \(\mathcal{T}\).

The projection $ \xi = \mathcal{P}(\eta)$ is obtained by solving the differential equations

\begin{align}
\dot{x}(t) &= \mathcal{H}_N(t)\,x(t), \qquad x(0)=x_0,\\
\boldsymbol{u}(t) &= \mu(t)-K_r(t)\bigl[x(t)-\alpha(t)\bigr],
\end{align}
where the regulator $K_r(t)$  is a time-varying feedback gain
designed to ensure that the trajectory $x(t)$ tracks the curve
\(\alpha(t)\) as closely as possible. Within this setting, any \(\eta\) is mapped onto the trajectory manifold and any \(\xi\) that is already in the trajectory manifold remains unaffected.
Once the constrained optimal control problem has been reformulated as the unconstrained optimization problem \eqref{eq:unconstrained}, Newton's method can be applied directly to the projected functional. The Newton method seeks a local minimum by applying an iterative procedure in the form,

\begin{equation}
    \boldsymbol{u}^{(k+1)} = \boldsymbol{u}^k + \nu^k ,
    \label{eq:newton_m}
\end{equation}
where, given a guess $\boldsymbol{u}^k$, the descent direction $\nu^k$ is obtained by computing the local quadratic approximation given by
\begin{equation}
\nu^k = \arg\min_{\nu} \left\{ \nabla g(\eta^k)^{T}\nu +\frac{1}{2}\nu^{T}\nabla^{2}g(\eta^k)\nu \right\}.
\end{equation}

In practice, however, the above equation may fail to admit a bounded solution and the Newton's method may not work. To solve this issue, a quasi-Newton method is used which consists of replacing the Hessian $\nabla^{2}g(\eta^k)$ by a suitable approximation $G^k>0$, a positive-definite matrix. Additionally, to avoid convergence problems, a damped quasi-Newton method is considered replacing \eqref{eq:newton_m} by
\begin{equation}
    \boldsymbol{u}^{(k+1)} = \boldsymbol{u}^k + \gamma^k\nu^k ,
    \label{eq:damp_newton_m}
\end{equation}
where the step size $\gamma^k\in(0,1]$  is chosen to ensure a sufficient decrease in the cost function. The choice of $\gamma^k$ is performed by the so-called Armijo rule (details in Refs. \cite{Shao2022}).

At each projected Newton iteration, the first and second-order Fréchet derivatives of the projected functional are approximated by locally linearizing the projected dynamics and constructing a quadratic expansion of the cost functional around the current trajectory. The local linearization is characterized by the matrices
\begin{equation}
    A^k(t)=\mathcal{H}^{(k)},
\end{equation}
and 
\begin{equation}
    B^k(t)=\left.\frac{\partial
\left(\mathcal{H}_N(t)\right)}
{\partial \boldsymbol{u}}
\right|_{\boldsymbol{u}^k}x^k,
\end{equation}
which are evaluated along the current projected trajectory. The first-order expansion of the projected cost functional requires the gradients

\begin{align}
q^k(t) &= \nabla_x g_b\!\left(x^k(t)\right), \\
r^k(t) &= \nabla_u g_a\!\left(\boldsymbol{u}^k(t)\right), \\
\pi^k &= \nabla_x J_0\!\left(x^k(T)\right),
\end{align}
where $q^k$, $r^k$, and $\pi^k$ denote the gradients of the control and terminal cost functions evaluated along the current trajectory. To construct the local quadratic approximation required by the Newton method, the corresponding second-order derivatives  of the state, control, and terminal cost functionals are defined as,

\begin{align}
Q^k(t) &= \nabla^2_{xx} g_b\!\left(x^k(t)\right), \\
R^k(t) &= \nabla^2_{uu} g_a\!\left(\boldsymbol{u}^k(t)\right), \\
\Pi^k &= \nabla^2_{xx} J_0\!\left(x^k(T)\right).
\end{align}

Combined with the linearized dynamics, these second-order derivatives define a local quadratic approximation of the projected optimal control problem, yielding an auxiliary linear-quadratic optimization problem posed on the tangent space of the trajectory manifold. The necessary optimality conditions yield a differential Riccati equation governing the Newton feedback gains. The corresponding solution is obtained by first integrating backward in time the coupled Riccati and adjoint equations,
\begin{align}
-\dot{P}&=(A^k)^{T}P+PA^k-K_o^{T}R^kK_o+Q^k ,\\
-\dot{p}&=(A^k-B^kK_o)^{T}p-K_o^{T}r^k+q^k,
\end{align}
with $P(T)=\Pi^k$ and $p(T)=\pi^k$. The optimal feedback gain and feedforward term are given by

\begin{align}
K_o&=(R^k)^{-1}(B^k)^{T}P,\\
v_o&=(R^k)^{-1}\left(( B^k) ^{T}p+r^k\right).
\label{eq:feedback_feedforward}
\end{align}

Having obtained the feedback gain $K_o(t)$ and the feedforward term $v_o(t)$, the Newton search direction is finally computed by forward integration of

\begin{align}
\dot{z}^k&=A^k z^k+B^k\nu^k,&z^k(0)&=0,\\
\nu^k&=-v_o-K_o z^k,
\end{align}
The optimized control is then updated according to \eqref{eq:damp_newton_m} and $z^k(t)\in X$ denotes the local approximation of the state update associated with the Newton correction. The directional derivative of the projected cost functional is calculated as

\begin{equation}
\begin{aligned}
Dg(\eta^k)
={}&
(\pi^k)^T z^k(T)
\\
&
+\int_0^T
\left[
q^k(t)z^k(t)
+
r^k(t)\nu^k(t)
\right]
\,dt,
\end{aligned}
\end{equation}
which is required both to determine the step size $\gamma^k$ through the Armijo backtracking line search and to assess the convergence of the Newton iterations. The projected Newton iterations are repeated until the value of $-Dg(\eta^k)$ becomes too small or the loss functional $J_0$ reaches a desired value.

The Newton search direction is obtained through the inversion of the Hessian block $R^k$; consequently, the conditioning of this matrix is critical to the numerical robustness of the algorithm (see Eq.~\eqref{eq:feedback_feedforward}). Excessively fine temporal discretizations may deteriorate the conditioning of the linear-quadratic subproblem, producing excessively large Newton updates and compromising the convergence of the projected iterations. For the systems considered in this work, stable convergence was achieved by selecting an appropriate control regularization matrix together with a suitable temporal discretization, thereby preserving a well-conditioned linear quadratic subproblem throughout the optimization.

For quantum gate optimization, all computational basis states are propagated simultaneously under the same collection of control fields. Consequently, a single set of optimized control functions is obtained, that simultaneously minimizes the gate error, limits the pulse fluence, and suppresses population leakage outside the computational subspace.

Unlike first-order methods such as Krotov, each PRONTO iteration requires the solution of a more complex step, the Riccati equation. Nevertheless, the resulting Newton updates substantially reduce the number of optimization iterations required to reach high-fidelity solutions, making the method particularly attractive for demanding quantum control applications involving multiple control fields. By combining trajectory projection with second-order optimization, it achieves fast local convergence while ensuring that all iterates remain dynamically feasible, thereby distinguishing it from conventional gradient-based optimal control algorithms.

\section{PINN for Quantum Optimal Control}
The basic idea of the PINNQOC method is to approximate the solutions of differential equations~\cite{RAISSI2019686} and to find the optimized controls using the NN. To accomplish such tasks, one needs to minimize a composite loss function, defined as
\begin{equation}
    L[\bm{\theta}_1,\bm{\theta}_2]= \sum_l c_l\cdot L_{l}[\bm{\theta}_1,\bm{\theta}_2],\label{eq:Loss}
\end{equation}
where $c_l$ are relative weights for each loss function and $\bm{\theta}_1$ and $\bm{\theta}_2$ represent the trainable weights and biases of the NN. The distinction between $\bm{\theta}_1$ and $\bm{\theta}_2$ lies in the fact that time-dependent states are functions of both $\bm{\theta}_1$ and $\bm{\theta}_2$, but external controls are exclusively related to $\bm{\theta}_2$, as shown below.
The first loss term implies that the NN must satisfy Eq.~\eqref{eq:scrodinger}; therefore,
\begin{widetext}
\begin{align}
    L_{1}[\bm{\theta}_1,\bm{\theta}_2] 
    &=\frac{1}{N_tN} \sum_{n=1}^{N_t}\sum_{j=1}^{N}
      \norm{  
        \frac{d|\psi_j^{[\bm{\theta}_1,\bm{\theta}_2]}(t_n)\rangle }{dt} 
        + i\left(H_0+ \sum_{m=1}^M u_m[\bm{\theta}_2,t_n] H_m\right)|\psi_j^{[\bm{\theta}_1,\bm{\theta}_2]} (t_n)\rangle}^{2},
    \label{eq:Lmodel}
\end{align}
\end{widetext}
where $|\psi_j^{[\bm{\theta}_1,\bm{\theta}_2]}(t_n)\rangle$ is the {\it j}-th quantum state that depends on $\bm{\theta}_1$ and $\bm{\theta}_2$, $N_t$ is the number of points used to numerically perform the time evolution of the Schrödinger equation, and $N$ denotes the number of initial states to be optimized simultaneously. 
The loss function $L_{1}[\bm{\theta_1,\bm{\theta_2}}]$ represents the dynamical equations at a set of points $\{t_n\}_{n=1}^{N_t}$ in the time domain, and minimization of this quantity ensures that the output of the NN follows the Schrödinger equation.
Since external controls are unknown, the solutions to this set of coupled differential equations are not unique. However, the uniqueness of solutions is ensured through the minimization of the boundary condition loss term, as follows 
\begin{widetext}
\begin{eqnarray}
    L_{2}[\bm{\theta_1},\bm{\theta_2}]&=&\frac{1}{N}\left( ||\ket{\Psi^{[\bm{\theta}_1,\bm{\theta_2}]}( 0)} -\ket{\Psi _{0}} ||^{2}\right.+\left.||\ket{\Psi^{[\bm{\theta}_1,\bm{\theta_2}]}(T)} -\ket{\Psi_{\text{tgt}}} ||^{2}\right).\label{eq:Ldata}
\end{eqnarray}
\end{widetext}

In this manner, the PINN framework requires the NN to simultaneously solve the differential equations and match the boundary conditions. To satisfy both of these requirements, the network must not only optimize $\bm{\theta_1}$ but also identify the external controls related to $\bm{\theta_2}$, to satisfy the boundary conditions. Modern automatic differentiation frameworks make this possible by tracking the computational graph~\cite{graph1} that includes the NN outputs associated with the solution of the Schr\"odinger equation and to the external controls. This ensures that the NN learns to approximate the solutions of the differential equations together with the optimization of external controls that minimize the boundary conditions $L_2[\bm{\theta_1},\bm{\theta_2}]$. To penalize the population of undesirable states, the following loss function must be included
\begin{equation}
\begin{split}
    &L_{3}[\bm{\theta_1},\bm{\theta_2}]=\frac{1}{N_tN_\phi} \\& \times \sum_{n=1}^{N_t}\sum_{j=1}^N\left\langle{\psi^{[\bm{\theta}_1,\bm{\theta_2}]} _j(t_n)}\right|P\left|\psi^{[\bm{\theta}_1,\bm{\theta_2}]} _j(t_n)\right\rangle ,
\end{split}
\label{eq:Lpop}
\end{equation}
where $P$ is the projection on the subspace to be avoided and $N_\phi$ is the number of states of this subspace.
Other terms might be included in the composite loss function, for example, probability conservation or regularization of the NN weights.

The algorithm works schematically as follows:

\begin{itemize}
    \item \textbf{Initialization:} Set the learning rate, the relative weights $c_l$ in Eq.~\eqref{eq:Loss}, the number of points $N_t$, and the maximum number of epochs.  Because $\ket{\psi_j^{[\bm{\theta}_1,\bm{\theta}_2]}(t_n)}$ is a complex function, we write $\ket{\psi_j^{[\bm{\theta}_1,\bm{\theta}_2]}(t_n)}=\ket{\psi_{j,R}^{[\bm{\theta}_1,\bm{\theta}_2]}(t_n)}+i\ket{\psi_{j,I}^{[\bm{\theta}_1,\bm{\theta}_2]}(t_n)}$, where the sub-indexes $R$ and $I$ denote the real and imaginary parts of each quantum state. Initialize a feedforward NN for $|\psi_{j,R}^{[\bm{\theta}_1,\bm{\theta_2}]}(t_n)\rangle$, $|\psi_{j,I}^{[\bm{\theta}_1,\bm{\theta_2}]}(t_n)\rangle$, and $ H_c[\bm{\theta}_2,t_n]$ with the trainable parameters $\bm{\theta_1}$ and $\bm{\theta_2}$.
    
    \item \textbf{Forward pass:} Evaluate the NN predictions related to $|\psi_{j,R}^{[\bm{\theta}_1,\bm{\theta_2}]}(t_n)\rangle$, $|\psi_{j,I}^{[\bm{\theta}_1,\bm{\theta_2}]}(t_n)\rangle$, and $ H_c[\bm{\theta}_2,t_n]$ for the inputs chosen.
    
    \item \textbf{Automatic differentiation and Backpropagation :} Calculate all necessary derivatives of $|\psi_{j,R}^{[\bm{\theta}_1,\bm{\theta_2}]}(t_n)\rangle$, $|\psi_{j,I}^{[\bm{\theta}_1,\bm{\theta_2}]}(t_n)\rangle$ and $ H_c[\bm{\theta}_2,t_n]=\sum_{m=1}^M u_m[\bm{\theta}_2,t_n] H_m$ as a function of time and evaluate Eq.~\eqref{eq:Loss}.  Calculate $\nabla_{\bm{\theta_1}} L[\bm{\theta_1,\bm{\theta_2}}]$ and $\nabla_{\bm{\theta_2}} L[\bm{\theta_1,\bm{\theta_2}}]$ as a function of parameters using automatic differentiation.

    \item \textbf{Optimization:}  With all the gradients computed, the optimizer will update both the network parameters $\bm{\theta_1}$ and $\bm{\theta_2}$.
    
    \item \textbf{Convergence:} Iterate the above steps until the composite loss function $L[\bm{\theta_1,\bm{\theta_2}}]$ 
    falls below a prescribed tolerance or the training reaches the maximum number of epochs.
\end{itemize}

\noindent
The outcome of this procedure is a trained NN for $\ket{\psi_j^{[\bm{\theta}_1,\bm{\theta}_2]}(t_n)}=\ket{\psi_{j,R}^{[\bm{\theta}_1,\bm{\theta}_2]}(t_n)}+i\ket{\psi_{j,I}^{[\bm{\theta}_1,\bm{\theta}_2]}(t_n)}$ approximating the system dynamics, and an estimate of external controls $u_m[\bm{\theta}_2,t_n]$ at the time grid.

We may also use hard constraints, as proposed in Ref.~\cite{Norambuena2023}, where the term related to the initial state in Eq.~\eqref{eq:Lpop} is replaced by the following modification:
\begin{equation}
    |\psi^{[\bm{\theta}_1,\bm{\theta_2}]}_j(t_n)\rangle=\left|\psi^0_j\right\rangle+g(t_n)\left|\varphi^{[\bm{\theta}_1,\bm{\theta_2}]} _j( t_n)\right\rangle ,\label{eq:transformed0}
\end{equation}
where $g(t)=1-\exp{(-t)}$. Equation \eqref{eq:transformed0} ensures that the initial condition is satisfied at t=0, but the loss functions in Eqs. \eqref{eq:Lmodel} and \eqref{eq:Ldata} must be modified accordingly as
\begin{widetext}
\begin{eqnarray}
    &&L^{\text{hc}}_{1}[\bm{\theta}_1,\bm{\theta}_2] 
    =\frac{1}{N_tN} \sum_{n=1}^{N_t}\sum_{j=1}^N
      \left\|\frac{g(t)d|\varphi_j^{[\bm{\theta}_1,\bm{\theta}_2]}(t_n)\rangle }{dt_n} +\exp{(-t_n)}|\varphi_j^{[\bm{\theta}_1,\bm{\theta}_2]}(t_n)\rangle\right.\nonumber\\
      &&\left. +i\left(H_0+ H_c[\bm{\theta}_2,t_n]\right)\left[\left|\psi^0_j\right\rangle+g(t)\left|\varphi^{[\bm{\theta}_1,\bm{\theta_2}]} _j( t_n)\right\rangle\right]\right\|^{2},
      \label{eq:Lmodel0}\\
      &&L^{\text{hc}}_{2}[\bm{\theta_1},\bm{\theta_2}]=\frac{1}{N}\sum_{k=1}^N\left( ||\left|\psi^0_j\right\rangle+g(T)\left|\varphi^{[\bm{\theta}_1,\bm{\theta_2}]} _j( T)\right\rangle -\ket{\psi_j^{\text{tgt}}} ||^{2}\right),\label{eq:Ldata0}
      \end{eqnarray}
\end{widetext}
where $\left|\varphi^{[\bm{\theta}_1,\bm{\theta_2}]} _j( t_n)\right\rangle$ are the quantities associated with the NN.

Furthermore, we propose a new type of hard constraint that automatically includes both boundary conditions at the initial and final time on the loss function related to the differential equations. To accomplish that, we perform the following transformation
\begin{widetext}
\begin{equation}
    |\psi^{[\bm{\theta}_1,\bm{\theta_2}]} _j(t_n)\rangle=f_T(t_n)\left|\psi^0_j\right\rangle+f_0(t_n)\left|\psi^{tgt} _j\right\rangle+f_0(t_n)f_T(t_n)\left|\varphi^{[\bm{\theta}_1,\bm{\theta_2}]} _j( t_n)\right\rangle ,\label{eq:transformed}
\end{equation}
\end{widetext}
where $f_T(t_n)$ and $f_0(t_n)$ are functions that satisfy the conditions $f_T(0)=f_0(T)=1$ and $f_T(T)=f_0(0)=0$. For simplicity, we choose $f_T(t_n)=(1-t_n/T)$ and $f_0(t_n)=t_n/T$ and the loss function given in Eq.~\eqref{eq:Lmodel} becomes
\begin{widetext}
\begin{eqnarray}
    &&L_{4}[\bm{\theta}_1,\bm{\theta}_2] 
    = \sum_{n=1}^{N_t}\sum_{j=1}^N
     \frac{1}{N_t} \left\|  
        f_0(t_n)f_T(t_n)\frac{d|\varphi_j^{[\bm{\theta}_1,\bm{\theta}_2]}(t_n)\rangle }{dt}+\left(\frac{f_T(t_n)-f_0(t_n)}{T}\right)|\varphi_j^{[\bm{\theta}_1,\bm{\theta}_2]}(t_n)\rangle+\frac{\left|\psi^{tgt} _j\right\rangle}{T}\right.\nonumber\\  
        &&-\frac{\left|\psi^{0} _j\right\rangle}{T}+\left.i\left(H_0+ H_c[\bm{\theta}_2,t_n]\right)\left[f_T(t_n)\left|\psi^0_j\right\rangle+f_0(t_n)\left|\psi^{tgt} _j\right\rangle+f_0(t_n)f_T(t_n)\left|\varphi^{[\bm{\theta}_1,\bm{\theta_2}]} _j( t_n)\right\rangle\right]\right\|^{2}.
    \label{eq:Lmodelhard}
\end{eqnarray}    
\end{widetext}
The numerical procedure to minimize the hard constraint loss function $L_{4}[\bm{\theta}_1,\bm{\theta}_2]$ is similar to the one described above, and this loss function replaces both the loss functions $L^{\text{hc}}_1[\bm{\theta}_1,\bm{\theta}_2]$ and $L^{\text{hc}}_2[\bm{\theta}_1,\bm{\theta}_2]$.

    
    
    
\section{Numerical Results}
In this section, we explore the numerical results of the three methods: Krotov, PRONTO, and PINN. The method that appeared to be more complicated to achieve good accuracy for the optimization of the control field is \newline PINNQOC, and different strategies are explored to improve the minimization of loss functions. Therefore, we first discuss this method and the strategies employed here to improve numerical accuracy.
\subsection{Models}
The simplest Hamiltonian for a leaking qubit is the three-level system, where the two lowest energy levels designate the qubit and the highest energy level is the leaking state.
The optimal control has already been successfully implemented using PRONTO~\cite{Shao2024} in the Hamiltonian below
\begin{eqnarray}
H=\sum_{i=0}^2E_{i}|i\rangle\langle i|+u_1(t)\left(\Omega_{01}|0\rangle\langle 1|+\right.\nonumber\\\left.\Omega_{02}|0\rangle\langle 2|+\Omega_{12}|1\rangle\langle 2|+\text{h.c.}\right),
\end{eqnarray}
where $E_0=0$, $E_1=1$ GHz, $E_2=5$ GHz, $\Omega_{0,1}=0.1$ GHz, $\Omega_{1,2}=0.5$ GHz, and $\Omega_{0,2}=0.3$ GHz. The physical platform modeled by this Hamiltonian is a three-level fluxonium qubit~\cite{Shao2024,PhysRevX.9.041041}.

In addition, we use a Hamiltonian that considers four levels~\cite{Norambuena2023}, where the two lowest energy levels represent the qubit, given by
\begin{eqnarray}
{H} &=&\sum_{i=1}^4E_{i}|i\rangle\langle i|+
 \frac{u_1(t)}{2} \left(|1\rangle\langle 3| + |1\rangle\langle 4|+ \text{h.c.}\right)\nonumber\\
&+& \frac{u_2(t)}{2} \left(|2\rangle\langle 3| - |2\rangle\langle 4|+ \text{h.c.}\right). 
\end{eqnarray}
This Hamiltonian is interesting for dealing with leakage because the coupling between the qubit states is indirect and the occupation of leaking states is mandatory in this case. We use the parameters that model a Nitrogen-Vacancy center that interacts with a Carbon-13~\cite{Awschalom2018}. These parameters are given by $E_1=E_2=E_3=0$, $E_4=6.79\;\text{a.u.}$~\cite{Norambuena2023,González_2022}.

\subsection{PINNQOC}

We start our numerical analysis by using PINNQOC to optimize the X-gate, without the penalty related to the leaking state. The architecture of the NN is a feedforward network composed of fully connected layers with 2 hidden layers of 128 neurons each. We use a learning rate of $10^{-3}$, the sine function as the activation function~\cite{NEURIPS2020_53c04118}, the input dimension is one (time $t$), and the output dimension is thirteen. 
 This output dimensionality accommodates a single control field, $u_1(t)$, alongside 12 ordinary differential equations (ODEs) that govern the dynamics of the quantum system.
 To accommodate the real-valued nature of standard NN architectures, the complex state vectors are decomposed into their respective real and imaginary components, such as $\ket{\psi_j(t)} = \ket{\psi_{j,R}(t)} + i\ket{\psi_{j,I}(t)}$. For a three-level system, each state vector is parameterized by the $6$ real-valued NN output units. To ensure the generation of a control field capable of implementing universal unitary operations rather than a specialized state-to-state transfer, the network is trained simultaneously across a complete set of initial basis configurations, $\{\ket{\psi_1^0}=\ket{0}, \ket{\psi_2^0}=\ket{1}\}$, yielding a total of $12$ real differential channels alongside the control field $u_1(t)$. The number of time steps is $N_t=500$ in this case.
 \begin{figure}[h]
    \centering
    \includegraphics[width=1.\linewidth]{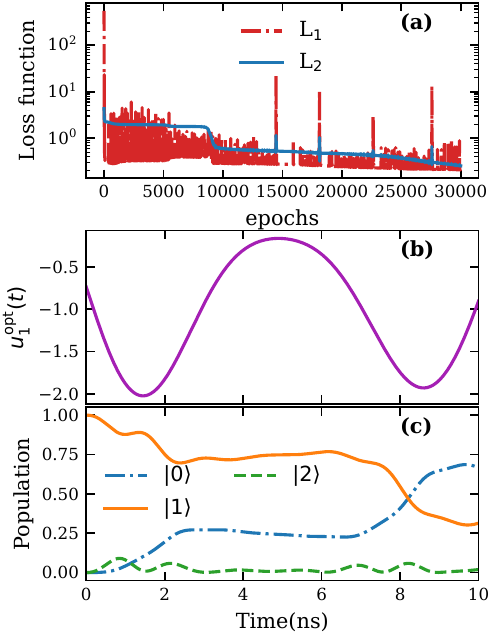}
    \caption{(a) Loss functions according to Eqs.~\eqref{eq:Lmodel} and \eqref{eq:Ldata} as a function of the number of epochs. (b) The optimized control field $u_1^{\textbf{opt}}(t)$ and (c) the populations for each state, obtained by driving the system with this optimized control, considering the initial state $|\psi(0)\rangle=|1\rangle$, as a function of time.
}\label{fig:Loss_functions_3levels_no_leakage}
\end{figure}
\begin{figure}[h]
    \centering
    \includegraphics[width=1.\linewidth]{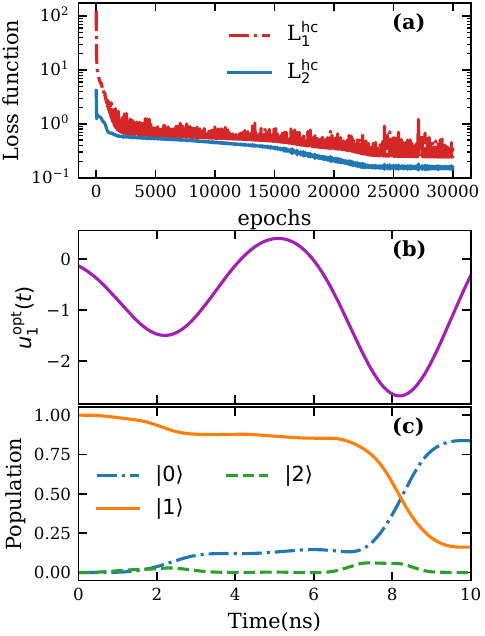}
    \caption{(a) Loss functions according to Eqs.~\eqref{eq:Lmodel0} and \eqref{eq:Ldata0} as a function of the number of epochs. (b) The optimized control field $u_1^{\textbf{opt}}(t)$ and (c) the populations for each state, obtained by driving the system with this optimized control, considering the initial state $|\psi(0)\rangle=|1\rangle$, as a function of time.
}\label{fig:Loss_functions_3levels_no_leakagehc}
\end{figure}
 Figure~\ref{fig:Loss_functions_3levels_no_leakage} (a) shows the loss functions described in Eqs. \eqref{eq:Lmodel} and \eqref{eq:Ldata} as a function of the number of epochs, where $c_1=c_2=1$ and $c_l=0$ for $l>2$ in Eq.\eqref{eq:Loss}. Both loss functions significantly oscillate and achieve a minimum value of the order of $2.1\times10^{-1}$ for $L_1$ and $2.5\times10^{-1}$ for $L_2$. The reason behind these severe fluctuations is that the PINNQOC framework must simultaneously optimize the control field and solve the system's dynamics. Consequently, the NN struggles to converge to a general solution for the ODEs because the differential equations are explicitly dependent on the external control; every update to the control field fundamentally alters the dynamics the network is attempting to learn. The optimized control field $u_1^{\textbf{opt}}(t)$ is plotted in Fig.~\ref{fig:Loss_functions_3levels_no_leakage} (b) as a function of time, and the populations obtained by driving the system with this optimized control are shown in Fig.~\ref{fig:Loss_functions_3levels_no_leakage} (c), considering the initial state $|\psi(0)\rangle=|1\rangle$. The population of the state $|0\rangle$ in the final time is 0.666, which is an unsatisfactory result for the X-gate. We discovered that the main reason for such poor performance of PINNQOC using this scheme is related to the imposition of achieving a quantum gate with a zero global phase (results not shown here). Although the results can be improved by letting the algorithm find the gate up to a global phase using the functional of Eq.~\eqref{eq:Jphi}, our goal is to find a scheme that works for a zero global phase (Eq.~\eqref{eq:gate_functional}).


To improve the optimization of the control field in this case, we use hard constraints as proposed in Ref. ~\cite{Norambuena2023}, where we replace Eqs.\eqref{eq:Lmodel} and \eqref{eq:Ldata} by Eqs. ~\eqref{eq:Lmodel0} and \eqref{eq:Ldata0}. The loss functions described in Eqs. ~\eqref{eq:Lmodel0} and \eqref{eq:Ldata0} as a function of the number of epochs are shown in Fig.~\ref{fig:Loss_functions_3levels_no_leakagehc} (a). Both loss functions also fluctuate significantly, but achieve a minimum value of the order of $2.4\times10^{-1}$ for $L^{\text{hc}}_1$ and $1.4\times10^{-1}$ for $L^{\text{hc}}_2$. 
The optimized control field $u_1^{\textbf{opt}}(t)$ and the populations, plotted in Fig.~\ref{fig:Loss_functions_3levels_no_leakagehc} (b) and (c) as a function of time, are different from those shown in Fig.~\ref{fig:Loss_functions_3levels_no_leakage} (b) and (c), and the population of the state $|0\rangle$ increases in the final time to 0.837.

\begin{figure}[ht]
    
    \includegraphics[width=1.\linewidth]{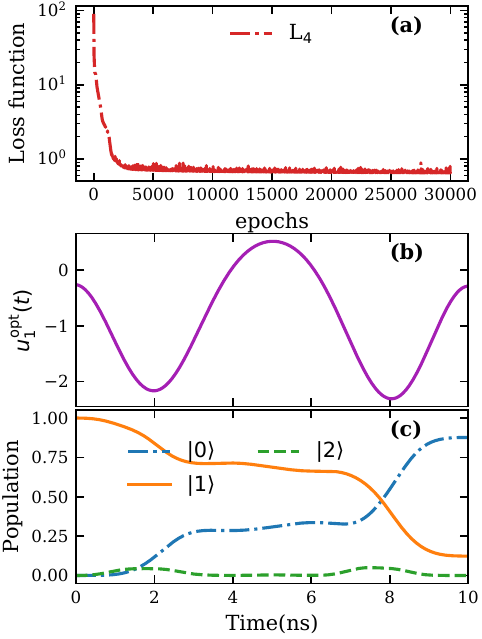}
    \caption{(a) Loss functions according to Eq.~\eqref{eq:Lmodelhard} as a function of the number of epochs. (b) The optimized control field $u_1^{\textbf{opt}}(t)$ and (c) the populations for each state, obtained by driving the system with this optimized control, considering the initial state $|\psi(0)\rangle=|1\rangle$, as a function of time.
}\label{fig:Loss_functions_3levels_no_leakageL4}
\end{figure}

In addition, we employ the two-boundary condition hard constraint, described by Eq. ~\eqref{eq:Lmodelhard}, to try to improve the optimization of the control field. In this case, where $c_4=1$ and $c_l=0$ for $l\neq4$ in Eq.\eqref{eq:Loss}. Figure~\ref{fig:Loss_functions_3levels_no_leakageL4} (a) shows the loss function described in Eq.  \eqref{eq:Lmodelhard} as a function of the number of epochs. The loss function is also very oscillatory and achieves a minimum value of the order of $6.45\times10^{-1}$. The optimized control field $u_1^{\textbf{opt}}(t)$ and the populations, plotted in Fig.~\ref{fig:Loss_functions_3levels_no_leakageL4} (b) and (c) as a function of time, are very similar to those shown in Fig.~\ref{fig:Loss_functions_3levels_no_leakagehc} (b) and (c), and the population of the state $|0\rangle$ increases in the final time to 0.876. The conclusion taken from Figs.~\ref{fig:Loss_functions_3levels_no_leakage}, ~\ref{fig:Loss_functions_3levels_no_leakagehc}, and ~\ref{fig:Loss_functions_3levels_no_leakageL4} is that the PINNQOC method faces problems during the minimization of loss functions and cannot achieve very high values of the quantum gate fidelity. We have also tested the same modifications for the loss functions for two-level and four-level Hamiltonians, and similar conclusions about the effectiveness of PINNQOC can be drawn from these systems (results not shown here).
On the other hand, PINNQOC is an effective tool for implementing quantum gates in the two-qubit system~\cite{lauten2025physicsinformedneuralnetworksgate}, although the number of control functions is larger than the one adopted here. Furthermore, it has been verified that the initialization weight plays an important role in minimizing the loss function~\cite{lauten2025physicsinformedneuralnetworksgate}. Based on this idea, we propose a procedure capable of solving the problem for a three-level system with high values of quantum fidelity. We first train the NN to reproduce the dynamics of the quantum system using a known control field, and we use the pretrained hyperparameters to initialize the minimization of the loss functions during the optimization of the control field. The control field for the pretraining of the NN is assumed to be a rapid oscillatory function $u_1(t)=\cos(20\pi t/T)$. We also need to force the normalization of the state vector in each epoch, according to the following equation
\begin{equation}
    |\psi^{[\bm{\theta}_1,\bm{\theta_2}]} _j(t_n)\rangle\rightarrow \frac{|\psi^{[\bm{\theta}_1,\bm{\theta_2}]} _j(t_n)\rangle}{|||\psi^{[\bm{\theta}_1,\bm{\theta_2}]} _j(t_n)\rangle||}.
\end{equation}

To perform pretraining, we need to minimize the following loss function
\begin{widetext}
\begin{eqnarray}
    L_{5}[\bm{\theta_1},\bm{\theta_2}]&=&\frac{1}{N_t N}\sum_{n=1}^{N_t}\sum_{j=1}^N||\ket{\psi^{[\bm{\theta}_1,\bm{\theta_2}]} _j(t_n)} -\ket{\psi^{[data]}_j(t_n)} ||^{2}\nonumber\\
    &+&\frac{1}{N_t M}\sum_{n=1}^{N_t}\sum_{m=1}^M |u_m^{[\bm{\theta_2]}}(t_n)-u_m^{0}(t_n)|^{2},\label{eq:Lpretraining}
\end{eqnarray}
\end{widetext}
where $\ket{\psi^{[data]}_j(t_n)}$ represents the value of the {\it j}-th state, obtained by exact evolution under the guess fields, at time $t_n$. The time evolved states can be evaluated through the partition of the evolution operator such as $\ket{\psi^{[data]}_j(t_n)}=\exp\left(-iH(t_{n-1})\Delta t\right)\ket{\psi^{[data]}_j(t_{n-1})}$, where $t_n=t_{n-1}+\Delta t$. The number of epochs for the pretraining is assumed to be $3000$.

\begin{figure}[ht]
    
    \includegraphics[width=1.\linewidth]{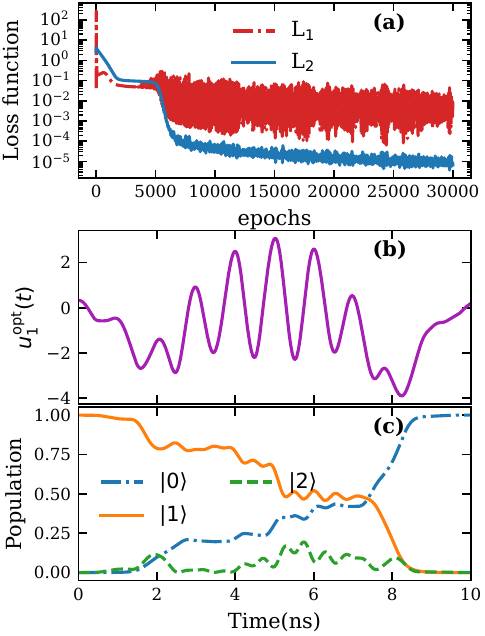}
    \caption{(a) Loss functions according to Eqs.~\eqref{eq:Lmodel} and \eqref{eq:Ldata} as a function of the number of epochs. (b) The optimized control field $u_1^{\textbf{opt}}(t)$ and (c) the populations for each state, obtained by driving the system with this optimized control, considering the initial state $|\psi(0)\rangle=|1\rangle$, as a function of time.
}\label{fig:Loss_functions_3levels_no_leakage_Fourier}
\end{figure}

NNs are known to exhibit a spectral bias, which means that they preferentially learn low-frequency functions before high-frequency ones~\cite{basri2020frequencybiasneuralnetworks}. This characteristic can hinder the accurate approximation of rapidly oscillating solutions, which are common in quantum dynamics. To overcome this limitation, we employ Fourier feature embeddings~\cite{tancik2020fourierfeaturesletnetworks}, which map the input variable into a higher-dimensional space spanned by sinusoidal functions. Specifically, the time variable $t$ is transformed according to
\begin{equation}
\chi(t)=\left[\cos(2\pi Bt),\sin(2\pi Bt)\right],
\end{equation}
where $B\in\mathbb{R}^{16\times1}$ is a vector of frequencies whose entries are independently sampled from a Gaussian distribution with standard deviation $\sigma=4$. This mapping produces a 32-dimensional feature vector, allowing the network to represent a broader range of frequency components, and thereby improve the approximation of highly oscillatory functions. It is worth-mentioning that these frequencies are not trainable parameters. Once sampled, they are fixed along the optimization procedure.
Taking into account the modifications described above, the results for the loss functions described in Eqs.~\eqref{eq:Lmodel} and \eqref{eq:Ldata} as a function of the number of epochs are shown in Fig.~\ref{fig:Loss_functions_3levels_no_leakage_Fourier} (a). Both loss functions achieve a minimum value of the order of $6.5\times10^{-5}$ for $L_1$ and $3.8\times10^{-6}$ for $L_2$, which are much smaller values than previously observed.
Notice that the optimized control field $u_1^{\textbf{opt}}(t)$ shown in Fig.~\ref{fig:Loss_functions_3levels_no_leakage_Fourier} (b) oscillates much more than those obtained without the Fourier feature (see Figs.~\ref{fig:Loss_functions_3levels_no_leakage} (b), \ref{fig:Loss_functions_3levels_no_leakagehc} (b), and \ref{fig:Loss_functions_3levels_no_leakageL4} (b)).

\begin{figure}[ht]
    \includegraphics[width=\linewidth]{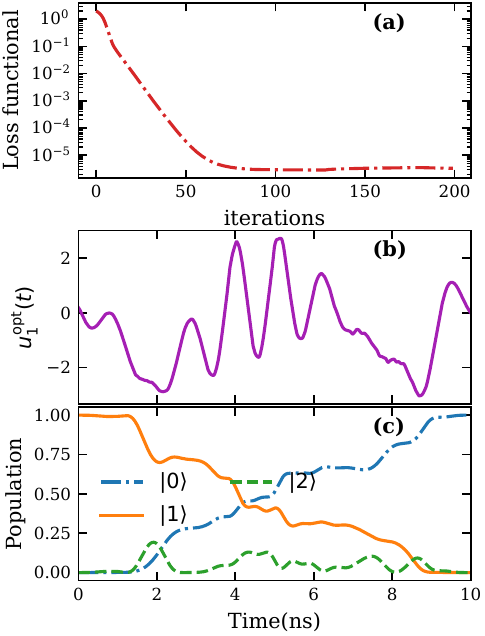}
    \caption{(a) Gate functional according to Eq.~\eqref{eq:gate_functional} as a function of the number of iterations for the Krotov method. (b) The optimized control field $u_1^{\textbf{opt}}(t)$ and (c) the populations for each state, obtained by driving the system with this optimized control, considering the initial state $|\psi(0)\rangle=|1\rangle$, as a function of time. A time grid with $N_t=1000$ collocation points and no leakage penalization was included.}
    \label{fig:Krotov_GateError}
\end{figure}
\begin{figure}[ht]
    \includegraphics[width=\linewidth]{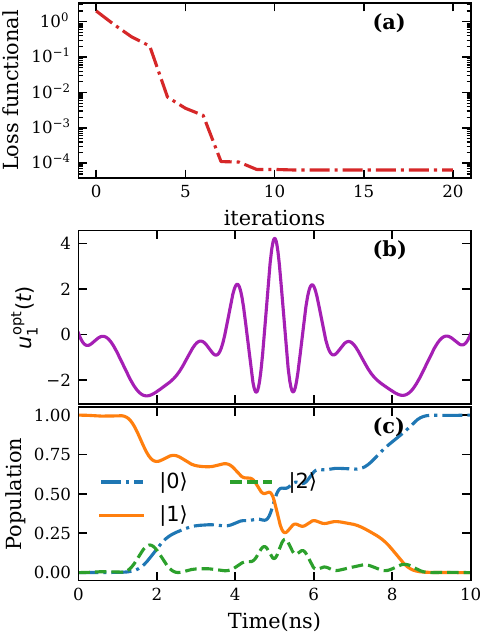}
    \caption{(a) Gate functional according to Eq.~\eqref{eq:gate_functional} as a function of the number of iterations for the PRONTO method. (b) The optimized control field $u_1^{\textbf{opt}}(t)$ and (c) the populations for each state, obtained by driving the system with this optimized control, considering the initial state $|\psi(0)\rangle=|1\rangle$, as a function of time. A time grid with $N_t=50000$ collocation points and no leakage penalization was included.}
    \label{fig:PRONTO_GateError}
\end{figure}
Figure \ref{fig:Loss_functions_3levels_no_leakage_Fourier} (c) shows the population of the state as a function of time, and the population of the state $|0\rangle$ increases to $0.99999$ in the final time. In summary, our new procedure is capable of solving the control problem with very high fidelity. The use of hard constraints within this new procedure does not bring any advantages (results not shown here); therefore, all the results presented hereafter will minimize the loss functions of Eqs.~\eqref{eq:Lmodel} and~ \eqref{eq:Ldata} by employing the pretraining, normalization in each epoch, and the inclusion of the Fourier features.

Figures~\ref{fig:Krotov_GateError} (a) and \ref{fig:PRONTO_GateError} (a) show the value of the functional $J_0$ defined in Eq.~\eqref{eq:gate_functional} as a function of the number of iterations of the algorithm for Krotov and PRONTO, respectively. We can observe that both Krotov and PRONTO require significantly fewer iterations than PINNQOC to achieve a steady value for the final-state loss functional. Krotov's method takes almost a hundred iterations, while PRONTO reaches the minimum value that the functional $J_0$ can achieve using this method after only 9 iterations. In principle, we could say that PRONTO far outperforms both Krotov and PINNQOC, and in turn, Krotov is superior to PINNQOC. But there are other factors we must take into account to decide which method to use, which will be analyzed in the next section. Regarding the difference in the optimized control fields obtained for each method, we can observe in Figs. \ref{fig:Loss_functions_3levels_no_leakage_Fourier} (b) and \ref{fig:Krotov_GateError} (b) that Krotov and PINNQOC reach control fields with similar amplitudes, but Krotov is likely to produce small and fast oscillations as the algorithm approaches the minimum, while the bounded frequency range inserted in the NN through Fourier features with Gaussian distribution allows the NN to reproduce the required oscillations, but prioritizes lower frequencies. In contrast, PRONTO seems to prioritize control fields with smaller frequencies and higher amplitudes, as we can see in Fig. \ref{fig:PRONTO_GateError} (b). Figures \ref{fig:Krotov_GateError} (c) and \ref{fig:PRONTO_GateError} (c) show the dynamics of the population of the three levels for the optimal fields obtained with Krotov and PRONTO, respectively, considering the initial state $|1\rangle$. Using Krotov and PRONTO, the population of the state $|0\rangle$ reaches a value of the order of $0.9999$, similar to the one obtained with PINNQOC. This suggests that the three methods are similarly effective in achieving the desired quantum gate operation. We will dive deeper into this fact in the next section.

\section{Performance Assessment and Discussion}

There are different features that we should analyze to compare these methods. We need to test how well the optimized fields perform the desired operation, and how costly the optimization procedure is for each method. To assess the effectiveness of the optimized Hamiltonian in jointly generating the quantum gate and avoiding leakage to higher energy levels, we use two quantities: the mean fidelity and the mean population outside the computational states. It is important to mention that even though the optimization procedure finishes with good fidelity, the quantum gate might not perform equally well for every initial state. 

To investigate the efficiency of quantum gate implementation, we consider the fidelity averaging over a set of $K = 500$ initial Haar random states $\ket{\psi^{\rm rnd}_j(0)}$. These computational states are uniformly distributed over the Bloch sphere. The mean fidelity can be defined as follows
\begin{equation}
    \mathcal{F}=\frac{1}{K}\sum_{j=1}^{K}\left|\bra{\psi^{\rm rnd}_j(T)} U_{\rm tgt} \ket{\psi^{\rm rnd}_j(0)}\right|^2,
\end{equation}
while the mean leakage is
\begin{equation}
    \ell=\frac{1}{K}\sum_{j=1}^{K}\frac{1}{T}\int_0^T
    \bra{\psi^{\rm rnd}_j(t)} P\ket{\psi^{\rm rnd}_j(t)} \, dt.
\end{equation}
Note that we are evaluating the fidelity at the final time and averaging the leakage over the entire time evolution, which is consistent with the goals of the optimization procedure. 

Regarding the computational cost of the implemented algorithm, we analyze the elapsed running time and the number of iterations it took for it to reach a desired tolerance for a common stopping criterion, defined as
\begin{equation}
    J_{0}\left[\left\{\ket{\psi _j( T)}\right\}\right]<10^{-3}
\end{equation}
as the stopping condition for all the optimization methods. 

Also, we should note that all numerical methods require a finite grid of time points, denoted by $N_t$. The control fields will only be evaluated on this grid and approximations are used for intermediate points. Insufficient grid resolution results in poor control fields efficiency. On the other hand, using a highly fine time grid, the optimization becomes highly costly, due to the sequential evaluation of the dynamics of the system.

To assess these quantities, we implement the first-order Krotov's method \cite{GoerzPackage}, the second-order PRONTO method \cite{Shao2024}, and our proposed PINNQOC framework on three- and four-level quantum systems. The resulting metrics are compiled in Tables I-VI. In the following subsections, we analyze these outcomes across three distinct dimensions: mean gate fidelity and mean leakage suppression, grid resolution constraints, and computational complexity. For all these simulations, we took a penalization factor for the populations (see the definition of the $g_b(\ket{\psi},t)$ term in Eq. \eqref{eq_func})) of $q=0.3$ for the three-level system and $q=0.2$ for the four-level system. In addition, for the Krotov's method, we did not use envelope functions ($S_j(t)=1 \, \forall \, j$), and we used the step widths $\lambda_j=1$ and $\lambda_j=10$ for the three- and four-level systems, respectively. For PRONTO, we took a weighting matrix $\mathcal{R}(t)=8 \times 10^{-4} \,\, \mathbb{1}$. Regarding the guess fields, we always used $u_1^0(t)=\cos \left(20 \pi t/T\right)$ and $T=10 \, \rm ns$ for the three-level system, and $u_1^0(t)=15(t/T) \exp\{-(3t/T)^2\}$, $u_2^0(t)=-u_1^0(t)$, and $T=3 \, \rm a.u.$ for the four-level system. All simulations were run on an AMD EPYC 7452 32-Core Processor, using no parallelization in order to test the algorithms in their basic form.

\subsection{Gate Fidelity and Leakage Suppression}
In the three-level fluxonium qubit system (Tables I-III), all three methods successfully achieve high mean fidelity gates exceeding the $99.8\%$ threshold, except the phase $S$ gate optimized via PINNQOC ($\mathcal{F} \approx 98.82\%$). In particular, for the $X$ gate (Table I), PINNQOC yields the highest mean fidelity of $99.972\%$, slightly outperforming Krotov ($99.915\%$) and PRONTO ($99.892\%$). This high fidelity is accompanied by a mean population leakage into the non-computational state $\ket{2}$ of approximately $\ell \approx 3.8\%$, which is comparable to Krotov ($2.7\%$) and PRONTO ($2.39\%$). This demonstrates that our proposed NN architecture, leveraging Fourier feature embeddings, successfully navigates the optimization landscape to resolve complex trajectories without sacrificing fidelity.

The physical challenge is significantly amplified in the four-level system (Tables IV-VI), where the coupling between the computational qubit states is indirect, rendering the population of the leaking levels mandatory during intermediate times. Under these constrained conditions, PINNQOC exhibits outstanding leakage suppression. For instance, in the phase $S$ gate (Table VI), PINNQOC reduces the mean leakage to just $7.39\%$, compared to $11.75\%$ for Krotov and $10.27\%$ for PRONTO. Similarly, for the $X$ gate (Table IV), PINNQOC limits leakage to $13.71\%$, whereas Krotov and PRONTO yield $24.71\%$ and $18.36\%$, respectively. This superior leakage mitigation highlights the capacity of the continuous neural representation to find smoother, more adiabatic-like control pathways that minimize the exposure of the quantum system to high-energy states.
\begin{table}[h]
    \centering
    \caption{Performance of the different methods to optimize a control field in a three-level system to achieve an $X$ gate while preventing leakage.}
    \label{tab:compar_3lev_X}
    \begin{tabular}{|c|c|c|c|c|c|}
        \hline
        Method &$N_t$& Iter.& Time (s) & $\mathcal{F}$&$\ell$ \\ 
        \hline
        Krotov & 100 & 393 & 6.64 & 0.99915 & 0.0270 \\
        PRONTO & 5000 & 70& 24.86 &0.99892 & 0.0239\\ 
        PINNQOC & 500 & 13501& 514.73 &0.99972 & 0.0381\\ 
        \hline
    \end{tabular}
\end{table}

\begin{table}[h]
    \centering
    \caption{Performance of the different methods to optimize a control field in a three-level system to achieve a Hadamard gate while preventing leakage.}
    \label{tab:compar_3lev_H}
    \begin{tabular}{|c|c|c|c|c|c|}
        \hline
        Method &$N_t$& Iter.& Time (s) & $\mathcal{F}$&$\ell$\\ 
        \hline
        Krotov & 100 & 109& 1.56&0.99805 & 0.0083 \\
        PRONTO & 5000 & 24& 24.23&0.99804 & 0.0080\\ 
        PINNQOC & 500 & 4489& 171.82&0.99819 & 0.0081\\ 
        \hline
    \end{tabular}
\end{table}

\begin{table}[h]
    \centering
    \caption{Performance of the different methods to optimize a control field in a three-level system to achieve a phase $S$ gate while preventing leakage.}
    \label{tab:compar_3lev_S}
    \begin{tabular}{|c|c|c|c|c|c|}
        \hline
        Method &$N_t$& Iter.& Time (s) & $\mathcal{F}$&$\ell$\\ 
        \hline
        Krotov & 1000& 2122& 112.56&0.99762 & 0.0294 \\
        PRONTO & 5000 & 207& 49.39&0.99428 & 0.0274\\ 
        PINNQOC & 500 & 15341& 577.28& 0.98823 & 0.0325\\ 
        \hline
    \end{tabular}
\end{table}

\begin{table}[h]
    \centering
    \caption{Performance of the different methods to optimize a control field in a four-level system to achieve an $X$ gate with a loss $L\left[\{\ket{\psi _{i}( T)}\}\right] < 10^{-3}$ while preventing leakage.}
    \label{tab:compar_4lev_X}
    \begin{tabular}{|c|c|c|c|c|c|}
        \hline
        Method& $N_t$& Iter.& Time (s) & $\mathcal{F}$&$\ell$\\ 
        \hline
        Krotov &100& 6780& 70.07& 0.99756& 0.2471\\ 
        PRONTO &5000& 120& 54.59& 0.99942 &0.1836\\ 
        PINNQOC & 500& 12571& 529.72& 0.99701 & 0.1371\\ 
        \hline
    \end{tabular}
\end{table}

\begin{table}[h]
    \centering
    \caption{Performance of the different methods to optimize a control field in a four-level system to achieve a Hadamard gate with a loss $L\left[\{\ket{\psi _{i}( T)}\}\right] < 10^{-3}$ while preventing leakage.}
    \label{tab:compar_4lev_H}
    \begin{tabular}{|c|c|c|c|c|c|}
        \hline
        Method& $N_t$& Iter.& Time (s) & $\mathcal{F}$&$\ell$\\ 
        \hline
        Krotov & 100& 2920& 27.53&0.99818 &0.1717 \\
        PRONTO & 5000& 120& 55.00& 0.99849& 0.1206\\
        PINNQOC & 500& 7973& 395.65&0.99861 &0.2699 \\
        \hline
    \end{tabular}
\end{table}

\begin{table}[h]
    \centering
    \caption{Performance of the different methods to optimize a control field in a four-level system to achieve a phase $S$ gate with a loss $L\left[\{\ket{\psi _{i}( T)}\}\right] < 10^{-3}$ while preventing leakage.}
    \label{tab:compar_4lev_S}
    \begin{tabular}{|c|c|c|c|c|c|}
        \hline
        Method& $N_t$& Iter.& Time (s) & $\mathcal{F}$&$\ell$\\ 
        \hline
        Krotov & 100& 1450& 14.63&0.99962 &0.1175 \\ 
        PRONTO & 5000& 32& 41.89&0.99908 &0.1027 \\ 
        PINNQOC & 500& 6549& 325.31&0.99241 &0.0739 \\ 
        \hline
    \end{tabular}
\end{table}

\subsection{Grid Resolution and Sensitivity ($N_t$)}

A key distinction between the methods lies in their sensitivity to the temporal grid resolution ($N_t$). Traditional solvers are highly coupled to the choice of time discretization; for example, PRONTO requires a highly refined grid ($N_t = 50000$ in Fig.~6, and $N_t = 5000$ in the tables) to guarantee the numerical stability of its underlying backward Riccati integration and trajectory projection steps. Reducing $N_t$ in PRONTO often leads to convergence failures or violation of dynamical feasibility.
On the other hand, Krotov proves to be highly robust at coarse resolutions, achieving convergence with only $N_t = 100$ in most scenarios, though it required $N_t = 1000$ for the $S$ gate in the three-level system.
Finally, PINNQOC achieves its high-fidelity gate configurations with a moderate grid size of $N_t = 500$. Because PINNs approximate the solution as a continuous differentiable function over the entire domain rather than discrete piecewise slices, they are less susceptible to discretization errors, offering a valuable trade-off between temporal resolution and gate performance.

\subsection{Algorithmic Convergence and Computational Costs}

Analyzing the elapsed execution time and the number of optimization iterations reveals a clear trade-off between the methods. Due to second-order Newton-step updates, PRONTO converges in the fewest iterations across all test cases (ranging from 24 to 207 iterations). However, each Newton iteration is computationally heavy, involving the linearization of the dynamics and backward-forward sweeps of the Riccati equations.
 Although Krotov requires significantly more iterations than PRONTO due to its first-order gradient updates ({\it e.g.}, 6780 iterations for the four-level $X$ gate), its updates are computationally inexpensive. Consequently, Krotov exhibits the fastest total elapsed times, often converging in just a few seconds ({\it e.g.}, $1.56$~s for the three-level Hadamard gate).
    PINNQOC requires the highest number of iterations (epochs) and the longest execution times (ranging from $171.82$~s to $577.28$~s). This is because the NN must simultaneously learn the global state dynamics and search for the optimal control parameters.

While this makes PINNQOC computationally heavier for these low-dimensional systems, this cost may be reversed for larger systems. Also, traditional solvers must undergo this entire computational effort from scratch if system parameters drift. In contrast, a trained PINN model can adapt to parameter drift almost instantly via transfer learning, or bypass sequential integration bottlenecks entirely in higher-dimensional Hilbert spaces by exploiting parallel tensor execution on modern GPU architectures.
\section{Conclusion}

In this work, we have systematically benchmarked the performance of PINNQOC against two established optimization solvers: the first-order Krotov method and the second-order PRONTO method. By testing these frameworks on a truncated three-level fluxonium qubit and a four-level Nitrogen-Vacancy center coupled to a Carbon-13 nuclear spin, we demonstrated that all three approaches are highly capable of suppressing population leakage into non-computational states while maximizing gate fidelity. While standard PINN formulations traditionally struggle with convergence and spectral bias in highly oscillatory multi-level dynamics, we successfully overcame these numerical bottlenecks by introducing Fourier feature embeddings, dynamic epoch normalization, and an informed pre-training routine. This allowed our enhanced PINNQOC to achieve high-fidelity gate operations, demonstrating parity with the physical accuracy of Krotov and PRONTO.  Although our benchmarking reveals that PINNQOC currently requires longer optimization times and more training epochs than traditional solvers, the NN paradigm holds immense potential for managing much larger quantum systems with superior speed in the future. Solvers like Krotov and PRONTO rely on the step-by-step numerical integration of differential equations at every optimization iteration, a process that scales poorly as the dimension of the Hilbert space expands. In contrast, NN architectures are composed of tensor operations that are natively parallelizable and can seamlessly exploit modern GPU and Tensor Processing Unit (TPU) hardware acceleration. Consequently, as quantum processors scale to high-dimensional multi-qubit spaces, the training of PINNs can be massively parallelized, potentially bypassing the sequential integration bottlenecks of conventional solvers.  Furthermore, while trajectory optimization methods like Krotov and PRONTO are strictly trajectory specific and must solve the entire optimization problem from scratch whenever a new target gate or physical parameter drift occurs, the NN paradigm offers unique pathways toward real-time adaptation. For instance, a pre-trained PINN can be rapidly recalibrated to compensate for parameter drift through transfer learning (warm-starting)~\cite{ash2020warm,perrone2018scalable}, requiring only a fraction of the optimization epochs compared to training from scratch. More importantly, this architecture can be extended into a parametric PINN by incorporating the Hamiltonian drift parameters or target gate specifications directly as additional network inputs~\cite{cho2024parameterized}. Once trained over a designated parameter space, such a meta-learning model~\cite{li2021fourierneuraloperatorparametric} could generalize to unseen drift scenarios, generating the updated optimal control fields almost instantaneously through a single forward pass. This real-time inference capability, combined with the prospect of fast parameter fine-tuning, positions PINNQOC as a highly promising candidate for closed-loop, adaptive quantum hardware control.  Ultimately, the new procedure demonstrated in this work positions physics guided machine learning as a robust paradigm that goes beyond acting as a mere alternative numerical solver. By bridging the inverse problem of hardware parameter learning with the direct problem of optimal pulse design, the PINN framework paves the way toward closed-loop, fully autonomous quantum processors. 
\begin{acknowledgments}
We acknowledge the support from Fundação de Amparo à Pesquisa do
Estado de São Paulo (FAPESP), project numbers (2025/08539-3, 2024/09015-5, 2024/22593-8, and 2024/09298-7). 
L.K.C. also acknowledges support from the National Institute of Science and Technology for Applied Quantum Computing through (CNPq 408884/2024-0).
\end{acknowledgments}

\bibliography{References}

\end{document}